# Model comparison of DBD-PA-induced body force in quiescent air and separated flow over NACA0015

Di Chen, Kengo Asada, Satoshi Sekimoto, Kozo Fujii
*Tokyo University of Science, Katsushika, Tokyo 125-8585, Japan*

Hiroyuki Nishida
*Tokyo University of Agriculture and Technology, Koganei, Tokyo 184-8588, Japan*

**Numerical simulations of plasma flows induced by dielectric barrier discharge plasma actuators (DBD-PA) are conducted with two different body-force models: Suzen-Huang (S-H) model and drift-diffusion (D-D) model. The induced flow generated in quiescent air over a flat plate in continuous actuation and the PA-based flow control effect with burst actuation in separated flow over NACA0015 are studied. In the comparative study, the body-force field and the induced velocity field are firstly investigated in the quiescent field to see the spatial difference and the temporal difference in a single discharge cycle. The D-D body force is computed with flush-mounted and bulge configuration of the exposed electrode, which is operated at the peak-to-peak AC voltage of 7kV and 10kV. The D-D models generate momentarily higher body force in the positive-going phase of the AC power, but activate smaller flow region than the S-H model with $Dc$ = 0.0117, which is given by the experiment beforehand at 7kV.[21] The local induced velocity of the D-D bulge case at 7kV measured in the downstream flow has the best agreement with the experimental result.[36] The maximum wall-parallel induced velocity in the S-H case with $Dc$ = 0.0117 is consistent with that in the experiment, however, the local induced velocity is relatively high with different flow structure. In the post-stall flow over the airfoil at the angle of attack of $12°$ and Reynolds number of 63000, the D-D bulge model at 7kV has approximately the same effect of leading-edge suction enhancement and reattachment promotion with the S-H model with $Dc$ = 0.16, which was previously proved to be sufficiently high to achieve the control performance.**

## I.  Nomenclature

| | | |
|---|---|---|
| a | = | speed of sound |
| BR | = | burst ratio; $T_{on}/T$ |
| $C_p$ | = | specific heat at constant pressure |
| D | = | coefficient of diffusion |
| $D_C$ | = | ratio between electrostatic body force and dynamic pressure in S-H model |
| e | = | total energy per unit volume |
| $e_c$ | = | elementary charge |
| $E_i$ | = | electric field vector |
| $f_i$ | = | body force vector (dimensional variable in body-force modelling) |
| $f^+$ | = | burst frequency |
| $f_{base}$ | = | base frequency of sine AC power wave |
| $f(t)$ | = | temporal body-force distribution in S-H model |
| $F^+$ | = | nondimensional burst frequency; $f^+ L_{ref}/U_{ref}$ |
| $F_{base}$ | = | nondimensional base frequency; $F_{base} L_{ref}/U_{ref}$ |
| $L_{ref}$ | = | characteristic length |
| $M$ | = | surface integral of momentum |
| $M_\infty$ | = | Mach number |
| $n$ | = | plasma density |
| $p$ | = | static pressure |
| Pr | = | Prandtl number |
| $q$ | = | heat flux vector |
| $Q_c, Q_{c,w}$ | = | charged density and surface charged density |
| $r_{ep}$ | = | recombination coefficients of electron-positive-ion |
| $r_{pn}$ | = | recombination coefficients of positive-negative-ion |
| Re | = | Reynolds number |
| $S_i$ | = | body force vector (non-dimensional variable in N-S equations) |
| $t$ | = | time |
| $T$ | = | burst wave period; $T_{on} + T_{off}$; $1/f^+$ |
| $T_{off}$ | = | period when sine wave switch is off during burst wave period |
| $T_{on}$ | = | period when sine wave switch is on during burst wave period |
| $T_{base}$ | = | period of sine AC power wave; $1/F_{base}$ |
| $u_i$ | = | velocity vector |
| $u_{max}$ | = | global maximum wall-parallel velocity |
| $u_l$ | = | local wall-parallel velocity |
| $U_{ref}$ | = | characteristic velocity |
| $V_{pp}$ | = | peak to peak voltage of AC power |
| $x_i$ | = | position vector |
| $\rho$ | = | air density |
| $\phi$ | = | external electric potential |
| $\gamma$ | = | ratio of specific heats |
| $\kappa$ | = | thermal conductivity |
| $\tau_{ik}$ | = | stress tensor |
| $\delta_{ij}$ | = | Kronecker delta |
| $\varepsilon_0, \varepsilon_r$ | = | vacuum and relative permittivity of dielectric layer |
| $\lambda_d$ | = | Debye length |
| $\alpha$ | = | coefficients of diffusion |
| $\eta$ | = | coefficients of ionization |
| $\delta_s$ | = | Dirac function |
| $\mu$ | = | charged particle mobility |
| $\sigma$ | = | surface charge density; $Q_{c,w}$ |

*Subscripts*

| | | |
|---|---|---|
| $inf$ | = | reference value |
| $e, p, n$ | = | electron, positive ion, and negative ion |

## II. Introduction

Active flow control utilizing plasma generators has raised much interest for its high availability and feasibility, dielectric barrier discharge plasma actuator (DBD-PA) is such a typical device for flow control, which only consists of two electrode and a dielectric layer between them, see Fig. 1(a). Due to the thin and light structure, DBD-PA can be attached on any flat or curved surfaces, corner or edge, where flow separation control is considered,[1]-[4] more importantly, without changing the original shape of an object. In recent 20 years, an increasing number of studies are conducted not only on separation control in Fig. 1(b,c), but also on noise reduction,[5] drag reduction[6] and flow transition delay.[7]

DBD-PA, in this paper a single asymmetric DBD-PA as we concerned, typically can generate a wall jet with the maximum velocity up to around 10m/s when the voltage and the base frequency of operating alternating current (AC) is 5-20kV (peak to peak) and 1-10kHz respectively. The ionization effect is largely determined by the applied voltage as well as the electric permittivity of dielectric layer. The mechanism of plasma-assisting flow control is believed as follows, the interaction of the ionized gas and neutral air result in an electrohydrodynamic body-force vector field coupling with the momentum transfer in the external flow at the downstream of the exposed electrode.[8]-[11]

To understand the flow phenomena with the plasma-induced body force, many research have made a great effort on the body-force modelling,[4],[12]-[14] specifically, the spatio-temporal distribution of the body force, which can be incorporated into high-fidelity flow simulation.[2],[7],[15],[16] The early models for body force field generated by a single DBD-PA, proposed by Massines et al.[12] and Roth et al.,[13] are one-dimensional (1-D) based on static formulation and does not account for the presence of the charged particles, therefore it barely fit to 2-D or 3-D applications. Furthermore, serval semiempirical models of 2-D plasma flow modelling, using linear,[14] exponential functions,[10],[11] and Gaussian distribution[4] of the spatial decay for the 2D body force component are reviewed by Corke et al.[17]

Experimental methods to investigate the body force production are largely employed with the flow measurement technics,[18] the body force vector can be determined in Navier-Stokes momentum equation with the measured velocity,[19],[20] using particle image velocimetry (PIV) and laser doppler velocity (LDV).

Simulation-assisted studies including simple analytical models are mainly discussed in current paper. Suzen and his colleagues[4] proposed the electrostatic model as the following equation derived from Enloe et al.'s,[10]

$$\boldsymbol{f}_i = Q_c \boldsymbol{E}_i = Q_c(-\nabla \phi), \qquad (1)$$

where $\boldsymbol{f}_i$ and $\boldsymbol{E}_i$ denotes body force vector and electric field vector, respectively. The force is contributed by two different parts: the external electric potential $\phi$ and the electric field created by the net charged density $Q_c$. The net body-force obtained by the analytical solution of Suzen-Huang (S-H) model is well validated by the experimental results,[21] however, the model ignored the complex plasma chemistry which leads to the highly unsteady forcing on the plasma flow.

In contrast, the charged-particle models associated with the first-principles fully-coupled approaches, considered the ordinary force diffusion, drift motion and Coulomb acceleration of electrons, and positive and negative ions, respectively. The so-called drift-diffusion (D-D) model was first developed for the physical modelling,[22]-[24] which focused on the electric-field effects on the charged particles. More recently, in spite of the time-consuming computing, D-D model was widely applied in the simulation of DBD-PA induced body-force field,[9],[25] however, few researches have input the D-D body force into flow simulation. Gaitonde et al.[26] conducted plasma-based stall control simulations with coupled first-principles approaches that largely reduced the complexity of the broad-spectrum problem. Nonetheless, the induced flow field of high temporal resolution during a single discharge cycle still remains unclear.

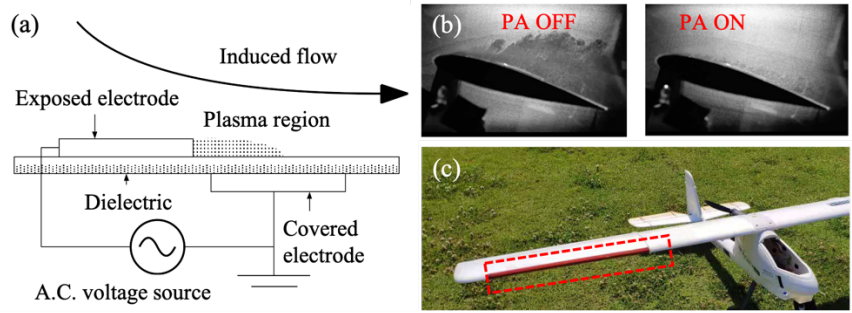

**Figure 1. (a) Sketch of DBD-PA, (b) wind tunnel test on airfoil, figures adapted with permission from Fujii (2018).[28] (c) DBD-PA attached on the leading edge of a small model plane, figures shot in a flight test.**

On the other hand, the analytical S-H models[4] and Shyy models[14] of low temporal resolution are well incorporated into the numerical flow simulations of Asada et al.[15],[16],[21] and Visbal et al.[2],[7] for their simplicity. The extra body force term in the Navier-Stokes equations include the parameter $D_c$ representing the scaling of the electrical to inertial forces. $D_c$ from another perspective describes the strength and the scale of body force, however, it is given empirically for each case considered. As to the transient accuracy of DBD-PA-induced flow, the S-H model utilizes the smooth and symmetrical sinusoidal function, while the D-D model shows the totally different discharge density in positive-going and negative-going phase of the AC power,[25] which is more realistic as it matches better with the experimental observation.[27]

To investigate the effect of the sophisticated D-D model on the high-resolution flow simulations, this research carries out a comparison between the D-D model and the S-H model, of which the latter one has been largely employed in the previous research.[15],[16],[21] The body-force model comparison is conducted in two typical flows: quiescent air over a flat plate and separated flow over NACA0015. Firstly, the body-force filed is showed in detail for the spatial and temporal difference in a single discharge cycle. Based on that, secondly, the simulations of continuous actuation are conducted with the two body-force models, which are compared in the global maximum wall-parallel velocity, the location of the maximum velocity, the local velocity measured in the downstream and the momentum integral in the near-field domain. Finally, the D-D model is applied in the high-resolution simulation of the separation control over the airfoil NACA0015 for the first time, which was largely studied before with the S-H model.[15],[16],[21],[28]

## III. Methodology

### III. A. Governing equations of fluid dynamics

The near-wall flow field driven by the plasma flow of DBD-PA in the quiescent air is described by the two-dimensional (2D) compressible Navier-Stokes equations, augmented by the terms representing the local forcing on the ionized region above a flat plate. The non-dimensional forms of the continuity, momentum, and energy equations, as well as the ideal gas equation are written as

$$\frac{\partial \rho}{\partial t} + \frac{\partial \rho \boldsymbol{u}_k}{\partial x_k} = 0, \quad (2)$$

$$\frac{\partial \rho \boldsymbol{u}_i}{\partial t} + \frac{\partial (\rho \boldsymbol{u}_i \boldsymbol{u}_k + p\boldsymbol{\delta}_{ik})}{\partial x_k} = \frac{1}{\text{Re}} \frac{\partial \boldsymbol{\tau}_{ik}}{\partial x_k} + D_c \boldsymbol{S}_i, \quad (3)$$

$$\frac{\partial e}{\partial t} + \frac{\partial ((e+p)\boldsymbol{u}_k)}{\partial x_k} = \frac{1}{\text{Re}} \left( \frac{\partial \boldsymbol{u}_l \boldsymbol{\tau}_{kl}}{\partial x_k} - \frac{1}{(\gamma-1)\text{PrM}_\infty^2} \frac{\partial \boldsymbol{q}_k}{\partial x_k} \right) + D_c \boldsymbol{S}_k \boldsymbol{u}_k \quad (4)$$

$$p = (\gamma - 1)(e - 0.5\rho \boldsymbol{u}_k \boldsymbol{u}_k) \quad (5)$$

where $\boldsymbol{x}_i$ represents the position vector, $\boldsymbol{u}_i$ the velocity vector, $\boldsymbol{q}_k$ the heat flux vector, $\rho$ the density, $p$ the static pressure, $e$ the total energy per unit volume, $\boldsymbol{\tau}_{ik}$ the stress tensor, $\boldsymbol{\delta}_{ij}$ the Kronecker delta, $\boldsymbol{S}_i$ the body force vector, $\gamma$ the ratio of specific heats, $t$ the time. In 2-D Cartesian system, the subscript $i$, $k$, $l$ denotes the wall-parallel and wall-normal direction. The above terms are all non-dimensional, the body force term $\boldsymbol{S}_i$ is normalized by

$$\boldsymbol{S}_i = \frac{1}{\rho_{\text{inf}} U_{\text{inf}}^2} \boldsymbol{f}_i, \quad (6)$$

where the net force $\boldsymbol{f}_i$ is computed in Eqn. (1). $D_c$ is the non-dimensional parameter, representing the ratio of the electrical force of the fluid to its inertial force, see details in the next section. In addition, the Reynolds number (Re), the Prandtl number (Pr), and the Mach number ($M_\infty$), are defined as follows,

$$\text{Re} = \frac{\rho_{\text{inf}} U_{\text{inf}} L_{\text{inf}}}{\mu_{\text{inf}}}, \text{Pr} = \frac{\mu_{\text{inf}} C_p}{\kappa_{\text{inf}}}, M_\infty = \frac{U_{\text{inf}}}{a_{\text{inf}}} \quad (7)$$

where $\mu$ is the viscosity, $L$ is the characteristic length, $U$ is the characteristic velocity, $a_{\text{inf}}$ is the speed of sound, $C_p$ is the specific heat at constant pressure, and $\kappa_{\text{inf}}$ is the thermal conductivity; the subscript inf indicates the reference values.

### III. B. Suzen-Huang body-force model

In this section, the spatial distribution of body force $\boldsymbol{S}_i$ in Eqn. (3) and (4) is presented by S-H model.[4] As we introduced from Eqn. (1), the body force vector is computed by multiplying the charge density $Q_c$ and electric field vector $\boldsymbol{E}_i$, which are solved in the Maxwell's equations of the external electric potential $\phi$ and the charged particle potential, respectively, as follows:

$$\nabla \cdot (\varepsilon_r \nabla \phi) = 0, \quad (8)$$

$$\nabla \cdot (\varepsilon_r \nabla Q_c) = \frac{Q_c}{\lambda_d^2}, \quad (9)$$

where $\varepsilon_r$ denotes the relative permittivity of the dielectric layer, and $\lambda_d$ denotes the Debye length. The boundary conditions of Eqn. (8) at the exposed electrode and Eqn. (9) on the wall above the covered electrode are shown in Fig. 2(b), can be written as

$$\phi(t) = \phi^{max} f(t), \tag{10}$$
$$Q_{c,w}(x,t) = Q_c^{max} G(x) f(t), \tag{11}$$

respectively, where $\phi^{max}$ and $Q_c^{max}$ are the maximum values of the external electric potential and the charge density, respectively. These two parameters controlling the strength of the plasma actuator's effects can be calibrated using available experimental data.[10]

The spatial distribution of the charged particles is given by a half Gaussian function $G(x)$ and the temporal variation of both Eqn. (10) and (11) can be a sine wave form $f(t) = \sin(2\pi\omega t)$, here $\omega$ is the base frequency of AC power, therefore $|f_i| \propto \sin^2(2\pi\omega t)$, as it is described in Fig. 2(a).

After substituting the solution of Eqn. (8) and (9) into Eqn. (1) and (6), we obtain the non-dimensional body force $S_i$, of which the magnitude is represented by multiplying the non-dimensional parameter $Dc$, here it is defined as

$$Dc = \frac{Q_c^{max} E^{max} L_{inf}}{\rho_{inf} U_{inf}^2}, \tag{12}$$

$E^{max} = (-\nabla\phi^{max})$ and $Q_c^{max}$ are the maximum magnitude of the electric field vector and the particle charge, respectively. To choose the appropriate $Dc$ value, Aono et al.[21] conducted many comparison cases between the experiments and the pre-computation of the S-H model.

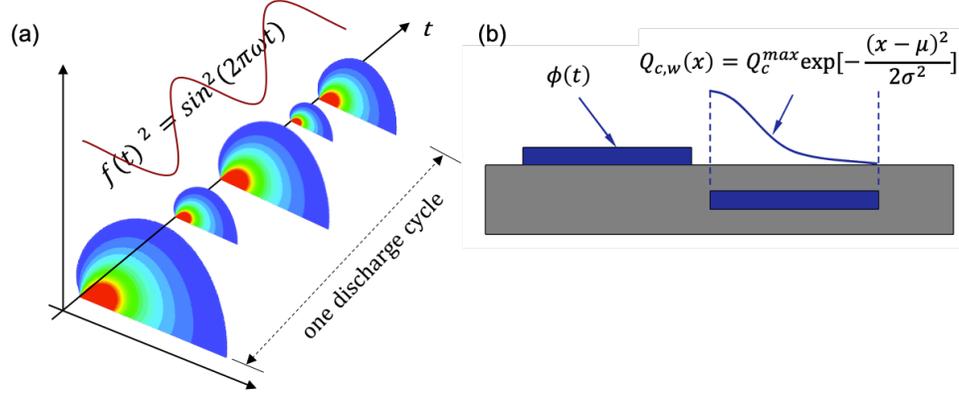

Figure 2. (a) Spatial and temporal distribution of body force in S-H model. (b) Boundary condition of charged particle in S-H model, following a half Gaussian distribution.[4]

### III. C. Two-dimensional drift-diffusion model

Notwithstanding the capability of the D-D model in 3D body force simulation by Nishida et al.,[25] the model comparison in 2D simulation is conducted for convenience and simplicity in this study. The electron, the positive ion, and the negative ion are considered with a basic plasma chemistry including electron impact ionization, attachment, and recombination. The time-dependent continuity equations for electron and ions with a D-D flux are coupled with Poisson equation. The governing equations are written as,

$$\frac{\partial n_e}{\partial t} + \nabla \cdot (-n_e \mu_e \mathbf{E} - D_e \nabla n_e) = (\alpha - \eta) n_e |v_e| - r_{ep} n_e n_p, \tag{13}$$

$$\frac{\partial n_p}{\partial t} + \nabla \cdot (n_p \mu_p \mathbf{E} - D_p \nabla n_p) = \alpha n_e |v_e| - r_{ep} n_e n_p - r_{pn} n_p n_n, \tag{14}$$

$$\frac{\partial n_n}{\partial t} + \nabla \cdot (-n_n \mu_n \mathbf{E} - D_n \nabla n_n) = \eta n_e |v_e| - r_{pn} n_p n_n, \tag{15}$$

$$\nabla \cdot (\varepsilon_r \mathbf{E}) = \frac{e_c}{\varepsilon_0}(n_p - n_e - n_n) + \frac{e_c}{\varepsilon_0} \sigma \delta_s, \tag{16}$$

where $n$ is the plasma density, $\mu$ is the charged particle mobility, and their subscript $e, p, n$ denote the electron, the positive ion, and the negative ion, $D$, $\alpha$ and $\eta$ are the coefficients of diffusion, ionization and attachment, $r_{ep}$ and $r_{pn}$ are the recombination coefficients of electron-positive-ion and positive-negative-ion, respectively. In Eqn. (16), $e_c$ is the elementary charge, $\sigma$ is the surface charge density similar to the boundary condition of $Q_{c,w}$ in Eqn. (11), here it is expressed by the Dirac function $\delta_s$. $\varepsilon_0$ and $\varepsilon_r$ are the vacuum and relative permittivity of the dielectric layer, respectively. The ionization and attachment coefficients and electron mobility are calculated by the BOLSIG[29] simulation software assuming the ambient gas is air ($N_2:O_2 = 0.8:0.2$). Other coefficients and parameters, partly shown in Fig. 3(b) keep the same with the work of Nishida et al..[25]

In the computation of electrohydrodynamic (EHD) force, as we call the body force, can be obtained by solving Eqn. (13-16), see details in the previous work of Nishida et al..[9],[25],[30] With the assumption in the previous studies, the body force is equal to the rate of momentum transfer per unit volume due to collisions,[23],[24],[30]

$$\boldsymbol{f} = e_c(n_p - n_e - n_n)\boldsymbol{E} - \left[\frac{D_p}{\mu_p}\nabla n_p + \frac{D_e}{\mu_e}\nabla n_e + \frac{D_n}{\mu_n}\nabla n_n\right], \quad (17)$$

on the right side, the first term of which is dominant in current condition, also correspond to that in Eqn. (1). Therefore, the unipolar region of the discharge plasma is the main region which contributes to the EHD force. Note that $D_c = 1$ when body force $\boldsymbol{f}$ in D-D model is normalized and substituted into N-S equations.

### III. D. Numerical approaches

Two typical flow are considered in the PA-based flow control. First, to investigate how strong the induced flow is produced, the flow field is assumed to be globally quiescent and laminar above a flat plate. Second, the separated flow over NACA0015 airfoil at the post-stall angle of attack of $12°$ is chosen to see the availability of D-D model. The Reynolds number is 63,000, the Mach number is 0.2, the specific heat ratio ($\gamma$) is 1.4, and the Prandtl number (Pr) is 0.72, assumed to be the same as those used in the previous simulation[16] and experiment setup.[21]

To solve the 2D Navier-Stokes equations in Eqn. (2)-(5), a compressible fluid analysis solver, LANS3D[31],[32] is employed. All the spatial derivatives are obtained with a sixth-order compact difference scheme.[33], besides MUSCL scheme is applied in the quiescent air before the steady state in which the starting vortex fades away. The low-order scheme with larger timestep and constant D-D force input makes for reducing the computational cost in the initial stage. With the time-varying D-D force input, the non-dimensional time increment normalized by the reference velocity and chord length, is set to $1 \times 10^{-5}$ in order to match the input frequency of the transient body force of the D-D model. Correspondingly, the Courant–Friedrichs–Lewy (CFL) number is less than approximately 1.0. Lower-upper symmetric alternating direction implicit and symmetric Gauss-Seidel (ADI-SGS)[34] method is utilized for time integration. Near the boundary, second-order explicit difference schemes are used because of the unavailability of high-order compact difference scheme. Tenth-order filtering[33] is applied with a filtering coefficient of 0.42. On the surface of the flat plate and the airfoil, no-slip and adiabatic conditions are imposed.

As to the body force computing in the D-D model, our current work largely follows the work of Nishida et al..[9],[25],[30] The drift terms in Eqn. (13)–(16) are evaluated by the upwind scheme using the MUSCL interpolation, and the diffusion terms are evaluated by the central difference scheme. The Poisson equation (16) is solved by the successive over-relaxation (SOR) method with the semi-implicit technique. The time integration is conducted by the implicit scheme, with the constant time increment of $2 \times 10^{-10}$ [s], which is finer than the adaptive value using CFL condition in[25],[30]. Due to the complexity of the charged-particle simulation, the timestep of D-D model is several orders smaller than that of computational fluid dynamics (CFD). Totally 4.5 discharge cycles are computed by the D-D model, of which 3 cycles are selected for the phase-averaging input. The input frequency of transient body force in a single discharge cycle is 1000, while the S-H model employs a sine-varying body force for the transient input, see Fig. 2(a). The time interval of the input D-D force matches the timestep in CFD simulation in real scale.

The computational grids for both D-D body force and CFD are listed in Table 1. In the flat plate simulation, the 2D computational domains and grids are shown in Fig. 3(a), the D-D force (red zone) has to be interpolated into the CFD grids. The minimum grid spacing is $0.0002L_{ref}$ ($L_{ref} = 0.1m$). The grid size keeps same with that in the S-H modelling of Aono et al..[21] The C-type grid is designed for NACA0015, the near and far field are shown in Fig. 3(b) and (c), respectively. The length of the computational region in the span direction is set to 0.2 times the chord length. Similarly, the D-D force is interpolated into the actuator zone (green), of which the minimum grid spacing is $0.0001L_{ref}$. The end of the exposed electrode of DBD-PA is placed at 5% chord length from the airfoil leading edge. Please note that the computation of D-D force relies on ultrafine grid according to the timestep. The sensitivity tests of grid resolution in both cases were well implemented in the previous studies.[21],[25]

Table 1. Grid density of the body-force computation and CFD in the generalized curvilinear coordinates

|  | $\xi$ | $\eta$ | $\zeta$ | Total points |
|---|---|---|---|---|
| D-D body force | 600 | 3 | 250 | 450,000 |
| Flat plate | 873 | 3 | 416 | 1,089,504 |
| Airfoil surface | 759 | 134 | 179 | 18,205,374 |
| Airfoil actuator | 149 | 134 | 111 | 2,216,226 |

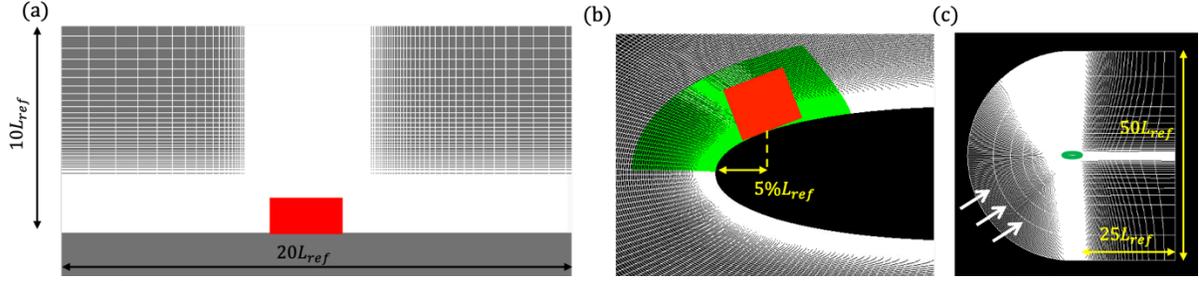

**Figure 3. Domain incorporation of the body-force computation (red) into CFD (background). (a) The flat plate, the yellow part is the body-force region. (b) NACA0015 airfoil, the red part is the body-force region of the bulge model, the green part is the actuator zone in CFD. (c) Computational domain of airfoil and schematic of inflow.**

### III. E. Computational setup of DBD-PA

Two configurations of DBD-PA in the D-D computational domains are shown in Fig. 4, called flush-mounted model and bulge model, see Fig. 4(a) and (b), respectively. As the thickness of the exposed electrode can be neglected in the flow, the major difference of the body force production between the two models comes from the thickness of the dielectric layer, and distance between the two electrodes and the length of the covered electrode. The flush-mounted model refers to the computational setup in S-H modelling, however, the D-D modelling makes the bulge model possible to copy the experiment setup.[21]

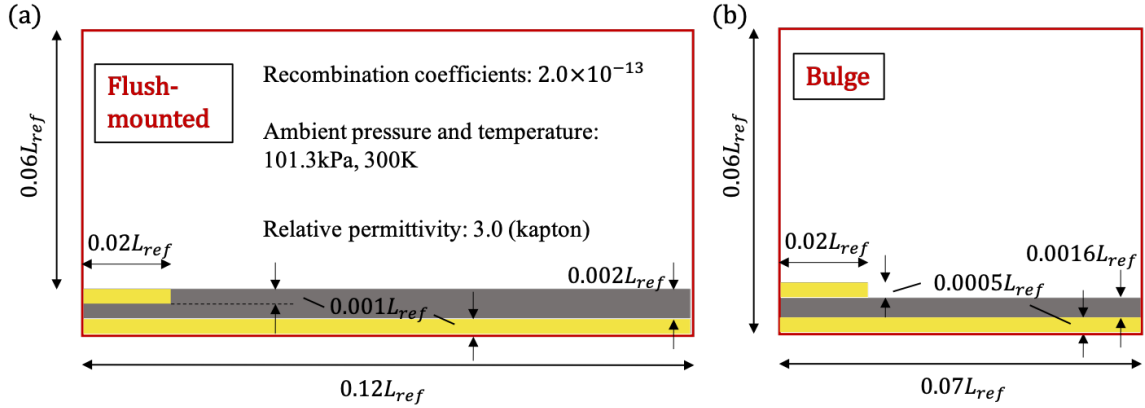

**Figure 4. Two types of PA configuration, and physical parameters for D-D computation ($L_{ref} = 0.1m$). Flush-mounted model (a) and bulge model (b), the setup refers to the CFD work and experiment for validation in Aono et al.,[21] respectively. Yellow denotes the exposed and covered electrode, grey denotes the dielectric layer.**

As it shows in Fig. 5, a sinusoidal form $0.5V_{pp}\sin(2\pi f_{base}t)$ of AC power is applied on the electrodes, of which the peak-to-peak amplitude ($V_{pp}$) is set to 7kV for the D-D force computation, and the base frequency ($f_{base}$) is 10kHz. As to the S-H model, $D_c$ in Eqn. (12) is set to 0.0117 corresponding to $V_{pp} = 7kV$, which is given by the experimental validation in the previous study.[21] The validation work was conducted associated with burst modulation of DBD-PA actuation. In the present study, continuous mode is applied to study the capability of body force production in the simulation of flat plate, where DBD-PA is permanently activated.

Burst mode is periodically activated in the separated flow over NACA0015 with the burst frequency $f^+ = 500$ [Hz] and the burst ratio $BR = 0.1$, which means a single burst contains two periods of the discharge cycle. Compared with the effect of continuous mode, the burst mode is proved to have higher suction peak at the leading edge with a proper $f^+$.[15],[16],[28],[35] Note that $F^+$ and $F_{base}$ are the normalized values of burst and base frequency, respectively, see Fig. 5.

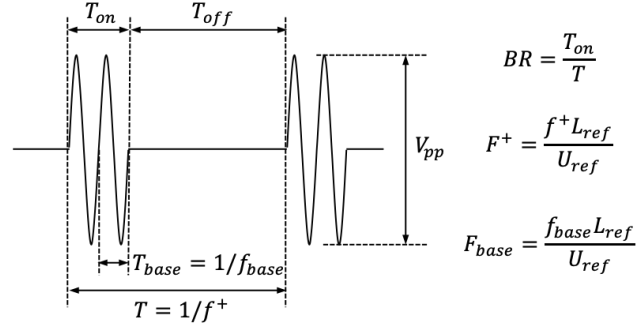

**Figure 5. Schematic diagram of a bursting wave in burst actuation mode.**

## IV. Results and discussions

### IV. A. Body force fields

Body force term on the right side of Eqn. (3) and (4) are computed by the D-D model and the S-H model. Fig. 6 shows two cycles of the EHD force (surface integral), which agrees with other simulation results.[24],[25] The S-H force is proportional to the function of $\sin(2\pi\omega t)$, and the force peak is quite close to those of the D-D force obtained by the flush-mounted model and bulge model at 7kV. The remarkable difference of the D-D forces between the positive-going and the negative-going voltage can be observed in this study. The push-push force is mostly produced in the positive-going phase, and the force reaches the peak simultaneously with the voltage peak. Moreover, in the positive-going phase the D-D force of the flush-mounted models shows some extremely high push-push force in one or two short moments. The D-D force is largely fluctuating in the second half of the negative-going phase, both the positive and negative force (push-push and push-pull) appear during the negative-going phase. However, the body force of S-H model is assumed to be positive all the time and symmetrical during negative- and positive-going phases, that may lead to the overestimated input power, see the time integral of S-H force.

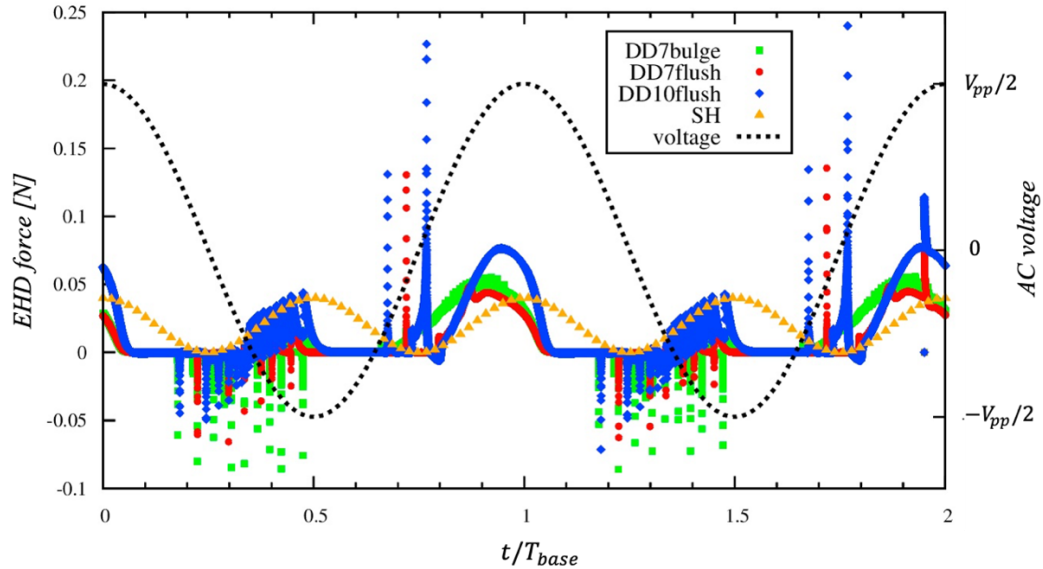

**Figure 6. Time history (two periods) of the dimensional area-integrated EHD force in the computing area, using the S-H model (SH), the bulge D-D model at 7kV (DD7bulge), and the flush-mounted D-D model at 7kV and 10kV (DD7flush & DD10flush). The corresponding AC voltage is also plotted.**

As the body force production in negative- and positive-going phase are quite different in Fig. 6, Fig. 7 further compares the distribution of the transient body force $S_x$ in the S-H model and the D-D models. $S_x$ is the non-dimensional horizontal body force (x-direction) per unit area expressed in Eqn. (6). $t/T_{base} = 0.5$ and $1.0$ are selected as the typical moments at the end of the negative-going phase and the end of positive-going phase, respectively. The upper row of Fig. 7 show much stronger body force of S-H model than those of D-D models in the first half discharge cycle, as well as the larger distribution area. However, the induced body force of S-H model is weaker than those of the D-D models in the second half discharge cycle, in spite of the larger induced area. The difference of force production between the flush-mounted model and the bulge model at the same 7kV is neglectable in Fig. 7, as the strength of the DBD-PA induced body force is largely determined by the input voltage. Enloe et al.[10] have indicated that the relation of the body force and the input voltage follow the power law of 3.35.

To show the spatial distribution of body force in detail, Fig. 8 further describes the time-averaged body-force fields near the end of the exposed electrode. The force profiles are plotted in a small streamwise region which is showed in Fig. 7 with the length of the electrode for reference. Both S-H model and D-D model in Fig. 8 show the decreasing trend of body force in the downstream direction, and the force of D-D model decays more rapidly. The maximum induced body force of D-D model is stronger than that of the S-H model.

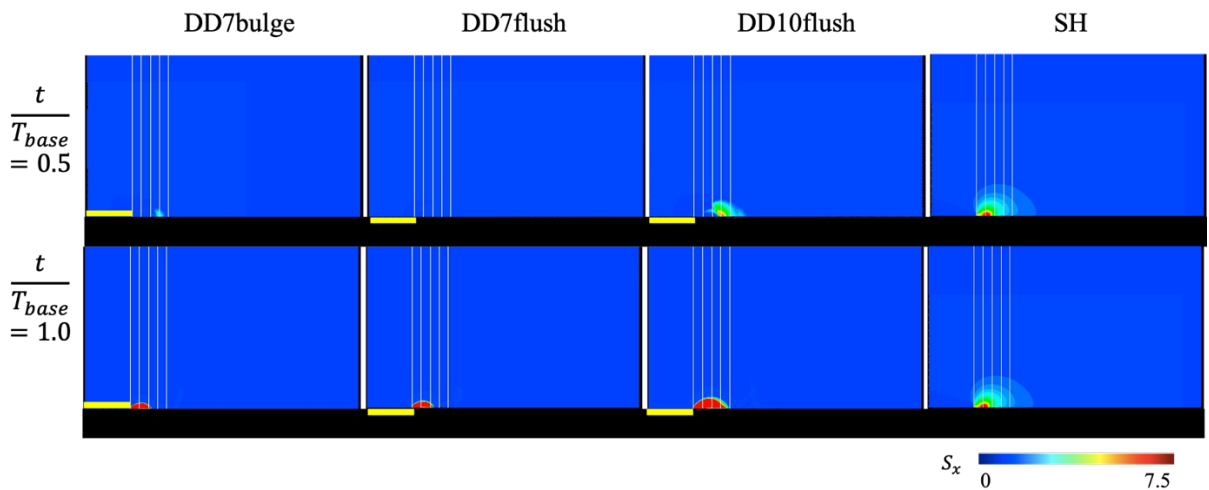

Figure 7. Transient horizontal body force at the voltage peaks obtained by S-H model and D-D models, the upper row shows $V = -V_{pp}/2$, the lower row shows $V = V_{pp}/2$. The figure tags are marked as the same with those in Fig. 6. The white lines denote the streamwise positions $x/L_{ref} = 0, 0.004, 0.008, 0.012, 0.016$ from left to right. $x/L_{ref} = 0$ is the end of the exposed electrode.

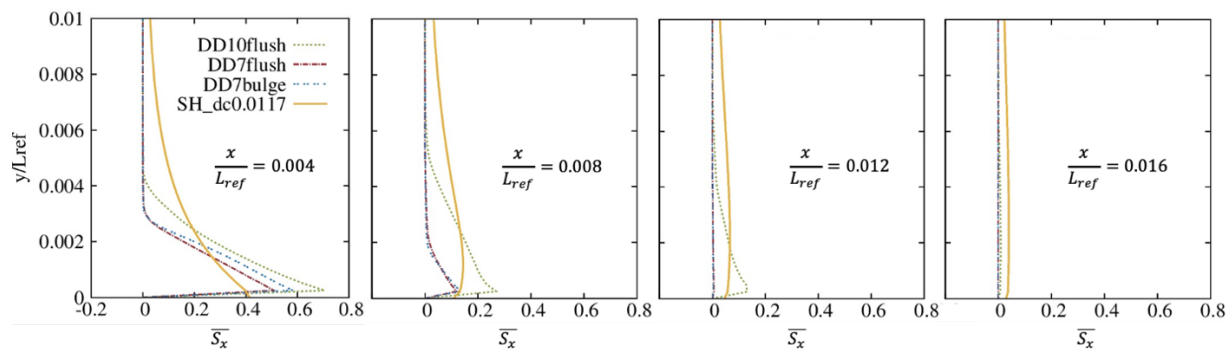

Figure 8. Profiles of time-averaged body force (horizontal) at different streamwise positions, which are showed in Fig. 7. $x/L_{ref} = 0$ is the end of the exposed electrode.

## IV. B. Quiescent fieds over a flat plate

It is quite straightforward to relate the body force production with the induced flow field in quiescent air. The flow field of S-H model corresponds to the case at 7kV, in which $Dc = 0.0117$ was determined in the previous research of Aono et al.[21] by a quantitative match between the computationally and experimentally measured maximum velocity in the wall-parallel flow. The previous validation was conducted with the burst modulation of actuation, however, the same value of $Dc = 0.0117$ in the present study is adopted for the continuous modulation. Fig. 9 shows the transient velocity fields in the wall-parallel direction in the continuous mode. The S-H model with $Dc = 0.0117$ is compared with the flush-mounted D-D model at 10kV, which has the strongest body force as Fig. 7 shows. The computational domain of D-D force is also showed with the black lines. The physical time is up to 0.4 seconds for both S-H and D-

D cases. The induced flow keeps enhancing at 0.4s in the S-H case, but that in the D-D case already becomes steady after 0.2s. The unsteady result of the S-H case indicates the insufficient computational time in the previous study,[21] which may cause the mismatch of the induced flow structure in the steady state during the continuous actuation.

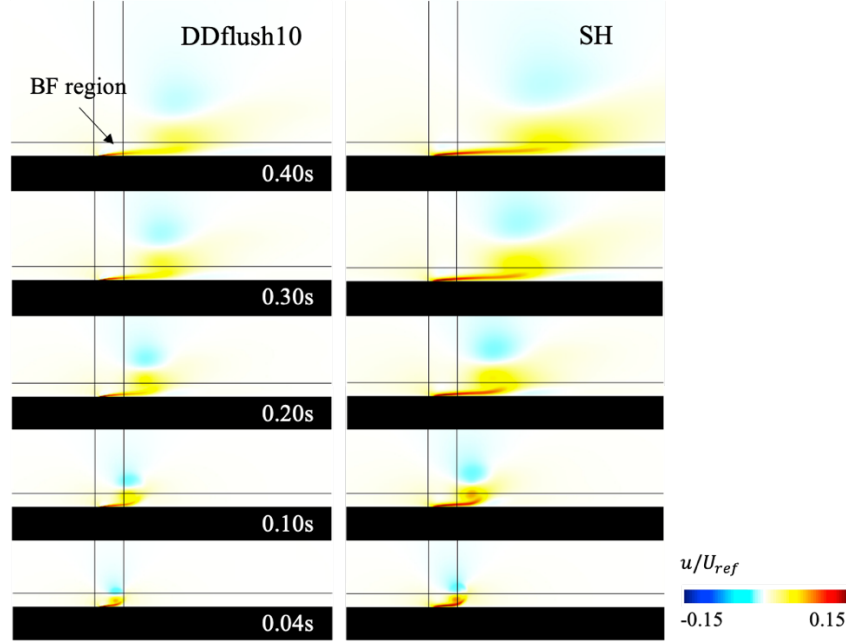

**Figure 9. Transient induced velocity fields from 0.08s to 0.4s. Here $u/U_{ref}$ is the non-dimensional velocity.**

The previous study of S-H model with the $Dc$ value of 0.0117 was conducted in the burst mode of actuation,[21] only with the time duration of 0.2s. The short time for the burst mode is probably enough to get a steady state of the induced flow, however, it is necessary to set a longer physical time of actuation in the present continuous mode to fully develop the induced flow field due to the stronger input power. Fig. 10 shows the time history of the surface integral of momentum,

$$M(t) = \int_{xs}^{xe} \int_{ys}^{ye} \frac{1}{2}\rho[u^*(x,y)]^2 dxdy, \tag{18}$$

where $u^* = u/U_{ref}$ is the non-dimensional velocity, xs, xe, ys, ye form a rectangular window ($\approx 1.13L_{ref} \times 0.75L_{ref}$) for calculating the momentum in it. To reach a steady state of the induced flow field of S-H model, the physical time is much long than 0.2s in Aono et al's[21] burst mode simulation, as well as 0.4s in Fig. 9. The time-averaged D-D force is used in the initial state before 0.8s, thus the larger timestep of the CFD simulation can be applied regardless of the input frequency of the D-D force in order to reduce the computational time.

The momentum integral obtained by the S-H model increases more rapidly than those by the D-D models in Fig. 10. It takes more time to get the constant momentum integral of the S-H case, which is also higher than those of the D-D cases. The momentum fluctuation of the D-D cases in Fig. 10 is caused by the time-varying D-D force input after 0.8s, the amplitude depends on the peak of body force showed in Fig. 6. As to the difference between the bulge D-D model and flush-mounted D-D model, the stronger body force input of the bulge model results in the higher power input (momentum integral). However, the effect of the AC voltage still plays the major role in the induced power.

Fig. 11 shows the flow fields at the physical time of 1.0s in the D-D and S-H simulations, which are confirmed as the steady state in the near field. The experimental result[36] is also included for comparison. The global maximum wall-parallel velocity ($u_{max}$) in the computational and experimental cases are measured in Fig. 11. Here $u_{max}$ is the averaged value in a single discharge cycle, it keeps constant from the very beginning of the actuation in both D-D and S-H model. Fig. 12 shows the maximum induced velocity against time during the discharge cycle, $u_{max}$ of the D-D models decrease till $3/4T_{base}$, then increases in the second half of positive-going phase of voltage ($3/4T_{base}$ to $T_{base}$), while $u_{max}$ of the S-H almost keeps constant.

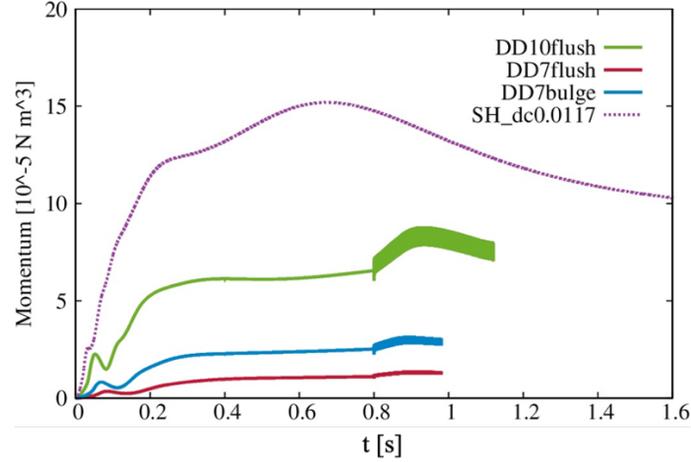

**Figure 10.** Time history of surface integral of momentum in a rectangular window of $1.13L_{ref} \times 0.75L_{ref}$.

The local maximum wall-parallel velocity ($u_l$) is also measured at $0.208L_{ref}$ downstream to confirm the steady state in the wake in Fig. 11. Both the flow structure and $u_l$ of the D-D bulge model at 7kV are close to the experimental results. The D-D flush-mounted model at 7kV and 10kV induces weaker and stronger flow, respectively, compared with the experimental result. The S-H model, however, produces much stronger induced flow with the $Dc$ value of 0.0117, which is determined by the $u_{max}$ measured in the experiment beforehand. In the present continuous mode of actuation, $u_{max}$ is constant during the actuation, and it is same with the value used in the burst mode of the previous study.[21] The magnitude of $u_{max}$ corresponding to $Dc = 0.0117$ in both continuous mode and burst mode are close to that of the experimental result at 7kV, see Fig. 12. However, it is supposed to be the larger induced region that the stronger flow is induced in S-H model, see the distribution of body force in Fig. 7.

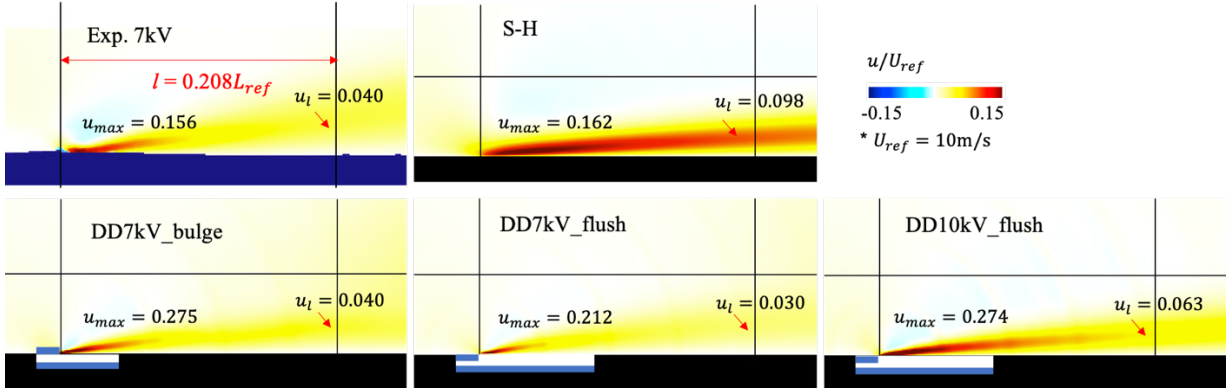

**Figure 11.** The transient induced flow field at 1.0s in S-H and D-D simulations, comparing with the experimental result from the dataset of Sekimoto et al.[36] $u_{max}$ is the maximum wall-parallel induced velocity in the entire domain, $u_l$ is the local maximum wall-parallel induced velocity at $0.208L_{ref}$ downstream. Both $u_{max}$ and $u_l$ are nondimensional.

To further investigate the near-field induced flow above the upper electrode, the locations of $u_{max}$ are searched during a single discharge cycle in the computational and experimental cases. The $u_{max}$ location in all the D-D cases are closer to the end of the upper electrode than those in S-H case and the experiment in Fig. 13. The $u_{max}$ location of the D-D bulge model at 7kV is further closer to the upper electrode at the end of the discharge cycle (end of the positive-going phase), meanwhile it brings higher $u_{max}$ compared with the other D-D cases. As to the S-H case, the vertical distance of $u_{max}$ location is approximately same with that of the experimental result, however, the parallel distance is over two times larger than the experimental one, that also indicates the larger induced region in S-H model.

To summarize the results above, comparing with the experimental data of $u_{max}$ and its location, S-H model has the same magnitude of $u_{max}$ in Fig.12 but the larger induced region in Fig. 13, which lead to the higher induced velocity in the downstream shown in Fig. 11. As to the D-D models, $u_{max}$ are higher than the experimental result, but less flow is induced, that probably results in the similar flow structure to the experiment in the wake.

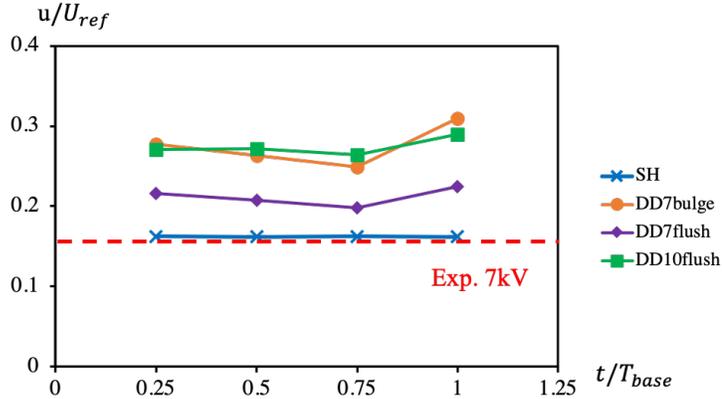

**Figure 12. The maximum wall-parallel induced velocity in a single discharge cycle.**

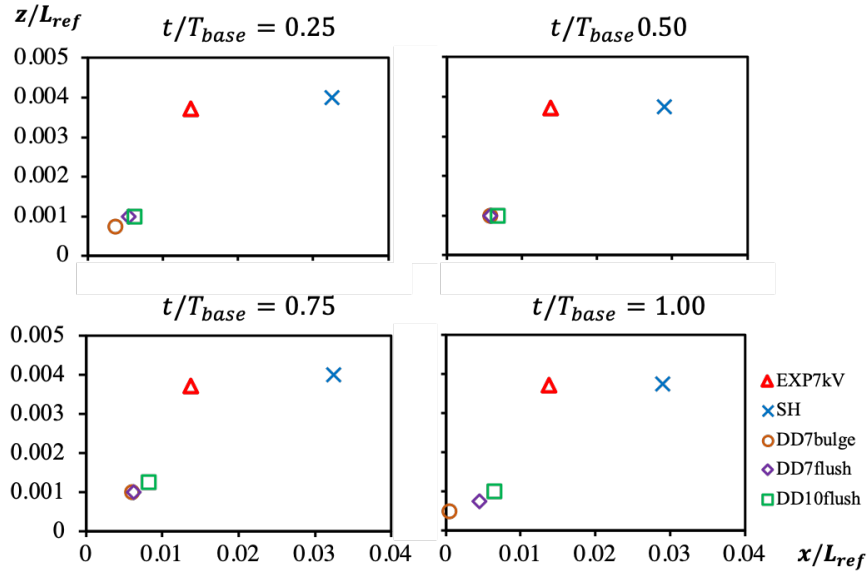

**Figure. 13. The location of the maximum wall-parallel induced velocity in a single discharge cycle.**

### IV. C. Separated flow fields over NACA0015

The simulations of flow control in the separated flow over airfoils based on the S-H model are largely implemented and well validated by experiments.[15],[16],[21],[37],[38] The actuators are numerically set with the burst mode, which periodically generates the induced flow. The effect of burst frequency on the flow control is of great interest in different separated flow fields, with a range of Reynolds number,[39] input voltage ($Dc$)[38] and different locations of the actuator.[38][39] In Subsection IV. B, the S-H model with $Dc = 0.0117$ is applied in the simulation of quiescent air over a flat plate, the maximum wall-parallel induced velocity in the continuous mode matches with the experimental result, moreover, the near-wall velocity profile in the burst mode of the previous study also shows the good agreement with the experiment.[21]

In this subsection, the comparative study between the D-D model and the S-H model moves to the separated flow over NACA0015 in order to see the flow control effect under the burst mode, where $F^+ = 5$ ($f_{burst} = 500Hz$). The D-D bulge model at 7kV is selected as the best fit to the experimental result at 7kV in the quiescent field for the study of separation control with the burst actuation. The S-H model with continuous mode in quiescent field shows the $u_{max}$ is about 1.6m/s for $Dc = 0.0117$. Many computational studies using S-H model utilize the sufficiently high $Dc$ in the separation control, $Dc = 0.04$ is widely used for the post-stall flow at $AoA = 12°$ and $Re = 63000$, corresponds to $u_{max} \approx 3.4$m/s.[37]-[39] $Dc = 0.1$ is also used in the same flow condition, corresponds to $u_{max} \in (3.4, 8.8)$.[38],[40]

$Dc = 0.016$ and $0.16$ are chosen for model comparison in the present study of separation control. Sato et al.[38] indicated that the lift coefficient of the case with $Dc = 0.16$ is approximately increased by 8% comparing with the case of $Dc = 0.01$. However, Sato et al.[37] also pointed out that the dependency of $Dc$ selection on the induced velocity is not so large when the momentum injection keeps the same. Here $Dc = 0.016$ is close to the value of 0.0117 which is used in the continuous mode simulation in the previous subsection, and $Dc = 0.16$ is 10 times higher to ensure the sufficient momentum injection.

Aerodynamic coefficients are shown in Fig. 14 in the chord-wise direction. 'DD7bulge' denotes the result obtained by the D-D bulge model at 7kV, 'NOACT' is the no control case. In the controlled cases, the time- and span-averaged pressure coefficient ($C_p$ in Fig. 14(a)) shows the plateau distribution of suction near the leading edge, which is contributed by the laminar separation bubble, found in both the PA-on phase and the PA-off phases. The suction plateau of the D-D case and the S-H case with $Dc = 0.16$ are approximately same, surpassing the value of the S-H case with $Dc = 0.016$. The $C_p$ curves indicate the lift increase in the controlled cases, while the skin-friction coefficient ($C_f$ in Fig. 14(b)) shows the reduction of pressure drag, as well as the separation region, see the grey region covered below zero. In the S-H case with $Dc = 0.016$, the attached region on the suction surface is slightly smaller and the reattachment point is delayed (farther away to the leading edge) compared with those of the other two cases, due to the less momentum injection in promoting the laminar-to-turbulence transition.

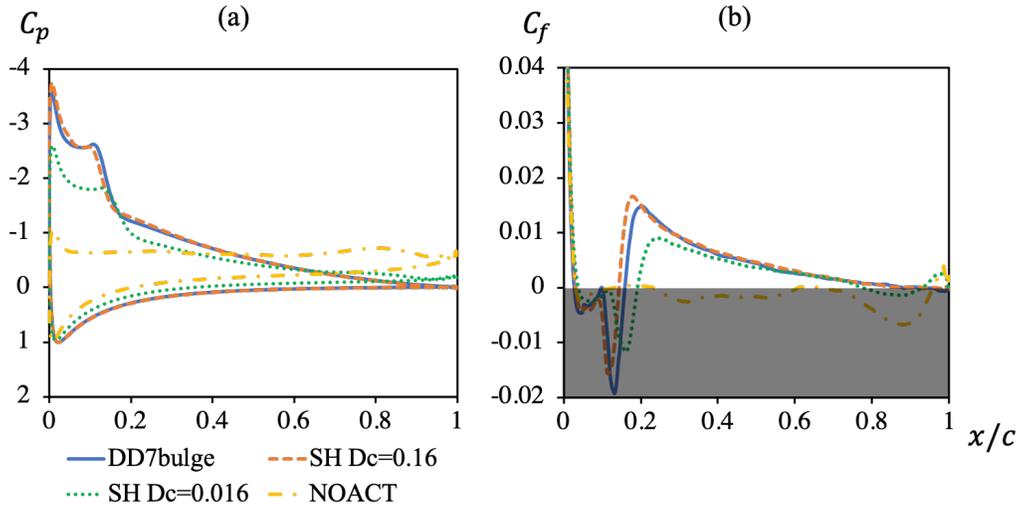

**Figure 14.** Time- and span-averaged pressure coefficient (a) and skin-friction coefficient (b) in non-dimensional time duration $tU_{ref}/c$ from 7.2 to 9.6. 'DD7bulge' the D-D bulge case at 7kV, 'SH $Dc$=0.16' and 'SH $Dc$=0.016' are the S-H cases with $Dc$ = 0.16 and 0.016, respectively, 'NOACT' is the no control case.

The instantaneous flow fields in Fig. 15 show the staged control effect at the nondimensional time 2.4, 4.8, 7.2 and 9.6, the flow may be regarded as the quasi-steady state between the latter two. The flow reattachment occurs earlier in the S-H case with $Dc = 0.16$ than the other two cases. The travelling spanwise vortices are observed in the D-D case and the S-H case with $Dc = 0.16$, result in the local oscillations of the pressure coefficient on the suction side mid-chord, however, this phenomenon does not occur in the S-H case with $Dc = 0.016$. The transient generation of spanwise vortices matches the frequency of the burst actuation, but finally becomes week in the quasi-steady flow, more details of the transient flow controlled by DBD-PA are studied previously by Asada et al.[16] with almost the same setup of actuator and flow condition.

The D-D model has nearly the same effect of separation control with the S-H model with the sufficiently high $Dc$. Compared with the no control case, all the three cases of the different body force model and the different $Dc$ value suppress the separation of the post-stall flow in the quasi-steady stage. However, the S-H model with $Dc = 0.016$ has the lower maximum induced velocity (see Fig. 12) but the higher injection momentum (see Fig. 10) than the D-D model in the continuous actuation, seems to be inferior to the D-D model in terms of enhancing the leading-edge suction in burst actuation. To accurately build the S-H model for both continuous and burst mode, $Dc$ and the momentum coefficient are probably not the only control factors. It is necessary to find the other dominant flow feature in the induced flow in the S-H model, which is remained as the future work.

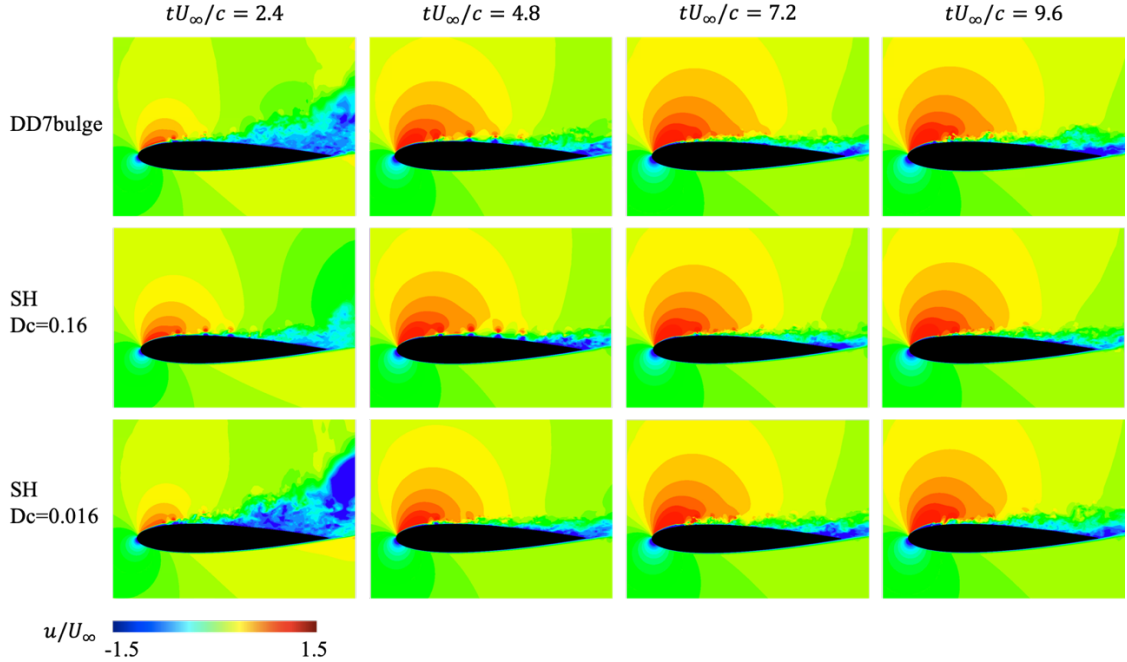

**Figure 15. Instantaneous flow fields obtained with the burst actuation $f_{burst} = 500 Hz$ ($F^+ = 5$). Colour contour is a chord-directional velocity.**

## V. Conclusion

In the present research, the simulations of the body force field induced by DBD-PA are conducted in Suzen-Huang model and drift-diffusion model. Based on that, the high-resolution CFD solver is employed to investigate the induced flow fields in the quiescent air over a flat plate with continuous actuation, as well as the separation control effect in the post-stall flow over NACA0015 with burst actuation. The D-D models of the two types of PA configuration generate most push-push body force in the positive-going phase of the AC voltage, including several extremely strong positive forces in very short moments. Conversely, the temporal distribution of the body force in S-H model is symmetrically set during the positive- and negative-going phases. The force peaks are approximately same between the D-D models at 7kV and the S-H model with $Dc = 0.0117$, which was given by the validation study of the quiescent field in burst actuation.[21] Therefore, the input power of the numerical actuator in the S-H model is supposed to be greater than those in the D-D models. This also directly leads to the increase of the momentum integral in the S-H model.

Compared with the experimental results of 7kV, the D-D models at the same AC voltage have the higher maximum induced velocity in the shorter distance from the end of the upper electrode, more importantly, generate almost the same magnitude of the local velocity in the wake. Moreover, the PA configuration has the minor effect on the maximum induced velocity, while the input voltage has the major effect on the induced flow structure. The best agreement is found between the S-H model and the experiment at 7kV[36] in the maximum induced velocity in wall-parallel direction above the flat plate, which is also the only factor in determining the value of $Dc$. The S-H model with $Dc = 0.0117$, however, results in the much stronger downstream flow than the D-D models and the experiment due to the larger induced flow region, speculated by the location of the maximum induced velocity in farther downstream.

As to the flow control in the separated flow with burst actuation, the D-D bulge model at 7kV is compared with the S-H model with $Dc = 0.016$ and 0.16, of which the higher $Dc$ value is widely proved to be sufficiently high to obtain the same performance with the experiment in flow control.[15],[16],[37]-[40] The D-D bulge model has almost the same effect of separation control with the S-H model with $Dc = 0.16$, in terms of increasing the lift peak at the leading edge, as well as moving the reattach point towards the leading edge. However, the S-H model with the lower $Dc$ is inferior to the D-D model in terms of enhancing the leading-edge suction. For future work, it may suggest modifying the S-H model to be more accurate for both continuous and burst actuation in various flow fields.

## Acknowledgments

The present study was supported in part by the Japan Society for the Promotion of Science (JSPS) through Grants-in-Aid for Scientific Research (No. 18H03816). The computation was supported by the supercomputer SX-ACE in Tohoku University. The authors also appreciated the valuable discussion with Prof. Marios Kotsonis from TU Delft.